\documentclass{aa}

\usepackage[version=3]{mhchem}
\usepackage{graphicx}
\usepackage{txfonts}
\usepackage{natbib}
\usepackage{color}
\usepackage{threeparttable}
\usepackage{array}

\begin{document}

   \title{\ce{O2} signature in thin and thick \ce{O2}-\ce{H2O} ices\thanks{Based on experiments conducted at the Max-Planck-Institut f\"ur Extraterrestrische Physik, Garching.
									   The raw spectra are available in electronic form at the CDS via anonymous ftp to cdsarc.u-strasbg.fr (130.79.128.5) 
									   or via http://cdsweb.u-strasbg.fr/cgi-bin/qcat?J/A+A}}

   \author{B. M\"uller\inst{1}
          \and
          B. M. Giuliano\inst{1}
          \and
          L. Bizzocchi\inst{1}
          \and
          A. I. Vasyunin\inst{1,2,3}
          \and
          P. Caselli\inst{1}}
        
   \institute{Max-Planck-Institut f\"ur Extraterrestrische Physik, Giessenbachstra\ss{}e 1, 85748 Garching, Germany\\
             \email{bmueller@mpe.mpg.de}
             \and
             Ural Federal University, Ekaterinburg, Russia
             \and
             Visiting Leading Researcher, Engineering Research Institute 'Ventspils International Radio Astronomy Centre' of Ventspils University of Applied Sciences, In\v{z}enieru 101, Ventspils LV-3601, Latvia}

   \date{Received - , 2018; Accepted - , 2018}

\abstract
{}
{In this paper we investigate the detectability of the molecular oxygen in icy dust grain mantles towards astronomical objects.}
{We present a systematic set of experiments with \ce{O2}-\ce{H2O} ice mixtures designed to disentangle 
how the molecular ratio affects the \ce{O2} signature in the mid- and near-infrared spectral regions. All the experiments were conducted in a closed-cycle helium cryostat coupled to a Fourier transform infrared 
spectrometer. The ice mixtures comprise varying thicknesses from 8 $\times$ 10$^{-3}$ to 3 $\mu$m. The absorption spectra of the \ce{O2}-\ce{H2O} mixtures are also compared to the one of pure water.
In addition, the possibility to detect the \ce{O2} in icy bodies and in the interstellar medium is discussed.}
{We are able to see the \ce{O2} feature at 1551 cm$^{-1}$ even for the most diluted mixture of \ce{H2O}:\ce{O2} = 9:1, comparable to a ratio of \ce{O2}/\ce{H2O} = 10 \% which has already been detected in situ in the coma of the comet 67P/Churyumov-Gerasimenko.
We provide an estimate for the detection of \ce{O2} with the future mission of the James Webb Space Telescope (JWST).}
{}


\maketitle

\section{Introduction}
Astrochemical models have always dedicated special attention to molecular oxygen.
With a cosmic abundance twice that of \ce{C}, atomic \ce{O} is the third most abundant element in space. In dense clouds, standard gas phase chemical models therefore suggest a comparable ratio of \ce{CO} and \ce{O2} after times $\geq$ 3 $\times$ 10$^{5}$ yr \citep[e.g.][]{Woodall2007}, where \ce{O2} is supposed to be formed especially via \ce{OH} + \ce{O} $\rightarrow$ \ce{O2} + \ce{H}.
The \ce{OH} here can be formed by the dissociative recombination of \ce{H3O+}, \ce{H3O+} + \ce{e-} $\rightarrow$ \ce{OH} + \ce{2H}.
However, observations with the Submillimeter Wave Astronomy Satellite (SWAS) by \citet{Goldsmith2000} towards Orion and with Odin by \citet{Larsson2007} towards $\rho$ Oph A showed a significant difference between model predictions and measurements.
The \ce{O2} abundances found were more than 100 times smaller than those predicted by models \citep{Goldsmith2000}.
Better agreement with observations can be obtained if freeze-out of \ce{O} atoms onto dust grains is taken into account in gas-grain chemical models \citep{Bergin2000,Viti2001}, with consequence surface production of \ce{H2O} and \ce{O2}, which may trap a significant fraction of oxygen, leaving only some atomic \ce{O} and \ce{CO} in the gas phase.
Observations conducted by \citet{Liseau2012} led to a \ce{O2} column density of N(\ce{O2}) = 5.5 $\times$ 10$^{15}$ cm$^{-2}$ with an upper limit of abundance of N(\ce{O2})/N(\ce{H2}) $\sim$ 5 $\times$ 10$^{-8}$ in warm gas (T $>$ 50 K) and to N(\ce{O2}) = 6 $\times$ 10$^{15}$ cm$^{-2}$ with a little higher abundance in cold gas (T $<$ 30 K).
\citet{Liseau2012} stated that detecting gas phase \ce{O2} might be so difficult because the \ce{O2} abundance is transient in $\rho$ Oph A and \ce{O2} is no longer detectable after $\sim$ 2 $\times$ 10$^{5}$ yr.
A relatively large amount of \ce{O2} has only been found with \textit{Herschel} in Orion as reported by \citet{Goldsmith2011}. This source is quite warm ($\geq$ 180 K), leading to a grain temperature of $\geq$ 100 K.
At this temperature the grains are warm enough to desorb \ce{H2O} ice and keep a large amount of oxygen with a big fraction in the form of \ce{O2} in the gas phase. Another explanation for the high \ce{O2} abundance found by \citeauthor{Goldsmith2000} is that low-velocity C-shocks might be responsible for the increase of molecular oxygen in the gas phase.

In the gas phase, the formation of \ce{O2} via
\[ \ce{OH} + \ce{O} \rightarrow \ce{O2} + \ce{H}, \]
has been discussed by for example, \citet{Davidsson1990} and \citet{Carty2005}.
According to \citet{Ioppolo2011}, \ce{O2} is formed in the solid state via
\[ \ce{O} + \ce{O} \rightarrow \ce{O2}. \]
Possible other formation paths for molecular oxygen as suggested by \citet{Ioppolo2011} such as
\[ \ce{HO2} + \ce{H} \rightarrow \ce{O2} + \ce{H2}, \]
are still under discussion.  With present \ce{CO} molecules, \citet{Ioppolo2011} reports the formation of \ce{CO2} rather than \ce{O2}.
Also, the reaction $\ce{HO2} + \ce{H} \rightarrow \ce{O2} + \ce{H2}$ can have a large activation barrier of 79.6 kJ mol$^{-1}$ \citep{Mousavipour2007} which makes it unlikely to proceed at low temperature in gas phase.
Reaction-diffusion competition, however, increases the probability of overcoming the reaction barrier on the ice surface \citep{Herbst2008}.
In the solid state, \ce{O2} is expected to form by the reaction of two O atoms, $\ce{O} + \ce{O} \rightarrow \ce{O2}$ \citep{Tielens1982,Hama2013}.
\citet{Dulieu2010} showed that the consecutive hydrogenation of \ce{O} in the solid phase,
\[ \ce{O} + \ce{H} \rightarrow \ce{OH}, \]
\[ \ce{OH} + \ce{H} \rightarrow \ce{H2O}, \]
is a very efficient process.
In addition to this path for water formation, molecular \ce{O2} is an important component for the formation of \ce{H2O} on grains via the production of \ce{H2O2} \citep{Ioppolo2008,Miyauchi2008},
\[ \ce{H} + \ce{O2} \rightarrow \ce{HO2} \xrightarrow{\text{\ce{H}}} \ce{H2O2} \xrightarrow{\text{\ce{H}}} \ce{H2O} + \ce{OH}, \]
\[ \ce{H} + \ce{OH} \rightarrow \ce{H2O}, \]
or the formation of \ce{O3} \citep{Jing2012},
\[ \ce{O} + \ce{O2} \rightarrow \ce{O3}, \]
\[ \ce{H} + \ce{O3} \rightarrow \ce{O2} + \ce{OH}, \]
\[ \ce{H2} + \ce{OH} \rightarrow \ce{H2O} + \ce{H}. \]
\citet{Cuppen2007} state that the reaction between \ce{H} and \ce{OH} represents the main route of \ce{H2O} formation in diffuse and translucent clouds as \ce{H} is abundant in the gas phase there.
In dense clouds, however,  \ce{H2} + \ce{OH} is the dominant path for \ce{H2O} formation while \ce{H} + \ce{H2O2} also contributes significantly.
\ce{OH} radicals formed in the reactions \ce{H} + \ce{O3} and \ce{H2O2} + \ce{H} can then also react with \ce{H} and \ce{H2} as experimentally examined by \citet{Oba2012}.
While effective \ce{H2O} formation upon H irradiation of solid \ce{O2} has been observed for temperatures of 12-28 K, \citet{Oba2009} showed that a co-deposition of \ce{H} and \ce{O2} on an \ce{Al} substrate leads to formation of compact amorphous \ce{H2O} ice, consistent with astronomical observations in molecular clouds, as well as \ce{H2O2} up to a temperature of 40 K.
\citet{Mokrane2009} and \citet{Romanzin2011} experimentally studied the reaction path of \ce{H} + \ce{O3} forming \ce{H2O}. Using dust grains as substrate, \citet{Mokrane2009} shows that the reaction of \ce{H} + \ce{O3} can efficiently form \ce{H2O}, while \citet{Romanzin2011} confirm this possible path working with a gold-coated copper substrate.

The detection of solid \ce{O2} in the infrared (IR) is impeded as the molecular vibration of \ce{O2} is infrared inactive. 
However, \citet{Ehrenfreund1992} have shown that, when \ce{O2} is embedded in a matrix, the interaction with neighbouring molecules allows for the detection of a feature assigned to the vibration mode at 1551 cm$^{-1}$ 
(6.45 $\mu$m). The band strength of this feature depends on the molecular composition of the matrix, in terms of the ratio between different components, and on the chemical polarity.
In their work, \citet{Ehrenfreund1992} explored the effect of different ice compositions on the line position, line width and band strenght of the \ce{O2} feature. The authors find that strongly polar environments, such as water, as well as weakly polar environments such as \ce{CO}, 
have no so strong effect on the \ce{O2} molecule, as the presence of a non-polar environment, such as \ce{CO2} molecules.

The search for molecular oxygen in solid form in astronomical objects using the \ce{O2} infrared vibration was attempted in the late 1990s. \citet{Ehrenfreund1998} observed a deeply embedded proto-stellar object, S140 in the S 140/L 124 complex, using the Short Wavelength 
Spectrometer (SWS) on-board the Infrared Space Observatory (ISO). Cold dense clouds were also observed with ISO-SWS and ground-based observations by \citet{Vandenbussche1999}. However, the detection of solid \ce{O2} appears to be a difficult task for these objects, 
and the mentioned studies resulted in the determination of upper limits.
Recently, molecular oxygen in a concentration of up to 10 \% relative to water has been found in the coma of the comet 67P/Churyumov-Gerasimenko, as \citet{Bieler2015} and \citet{Mousis2016} have recently reported.
Successive work of \citet{Keeney2017} comparing line of sight measurements of \ce{O2} conducted with the Alice far-ultraviolet spectrograph on Rosetta with in situ measurements of the Rosetta Orbiter Spectrometer for Ion and Neutral Analysis (ROSINA) hint to even higher abundances of 11-68 \% with a mean value of 25 \%, although the comparison is not easily done.
This detection opens new scenarios and raises interest in the possibility of detecting molecular oxygen in cometary and interstellar ice.
Recent work by \citet{Taquet2016} show that solid \ce{O2}, with a molecular ratio to water ice similar to those measured in comet 67P, can be produced in molecular clouds with relatively low H to O abundance ratio in the gas phase, high densities ($\geq$ 10$^5$ cm$^{-3}$) and dust temperature around 20 K, higher than that typically measured in interstellar dark clouds (10-15 K).

In this paper, we implement new laboratory experiments based on the work of \citet{Breukers1991} as presented in \citet{Ehrenfreund1992}, where the 6.45 $\mu$m molecular oxygen feature was measured in an ice mixture of \ce{H2O}:\ce{CO}:\ce{CO2}:\ce{O2} $=$ 2:2:0.5:1 with unknown thickness. 
On the basis of this experiment, we have extended our analysis to different molecular composition ratios, as well as to different thicknesses. We have limited our analysis to the ice mixtures composed of \ce{H2O} and \ce{O2} to study solely the impact of the water matrix on the imbedded \ce{O2} feature, even though this molecular composition is not necessarely representative of the average composition of astrophysical ice.
Thus, we put strong emphasis on the dependence of the ice composition on the \ce{O2} band strenght.
Further studies with different mixtures, inclusive of \ce{CO} and \ce{CO2} at a 20-25 \% level, as found in interstellar ices \citep[e.g.][]{Boogert2015} will be presented in a future paper. 
This paper is structured as followed: We describe in Section 2 the methods and the setup used in the experiment. In Section 3 we present the experimental results and in Section 4 we discuss our measurements. Section 5 is dedicated to the astrophysical implications of
our experiments.

\section{Methods}
The experiments were conducted in the recently developed cryogenic laboratory at the Center for Astrochemical Studies (Max Planck Institute for Extraterrestrial Physics) in Garching (Germany). The experimental setup is composed by 
a high power cryocooler, purchased from Advanced Research Systems (ARS), hosted in the sample compartment of a high resolution Bruker IFS 120/5HR Fourier Transform Infrared (FTIR) spectrometer. 

The cryocooler consists of a cold finger situated into a vacuum chamber, with a final vacuum of 10$^{-7}$ mbar. A sample holder (model SHO-1A, also from ARS) is placed at the end of the cold finger, suited to be cooled down to a minimum temperature of 
4.2 $\mathrm{K}$. The temperature is controlled by a Lake Shore Cryotronics (model 335) temperature controller, equipped with DT-670 Silicon Diodes sensors. The measured lowest temperature reached by the sample holder is 5 $\mathrm{K}$. 
The sample holder can be complemented with substrates of different materials, on top of which the ices can be formed by condensation from the gas phase. The substrate used for this set of 
experiments is made of potassium bromide (\ce{KBr}), which offers, with a transmission of $>90 \%$ through the whole observed spectral range, the best optical properties in the desired frequency range. The appropriate gas mixture is introduced into the cryocooler by an expansion through a 6 mm
stainless steel pipe, attached to a gas reservoir. The exit of the pipe is placed at a distance of approximately 2 cm from the sample holder, facing the substrate. In this set of experiments the gas mixtures were deposited onto the substrate kept at a temperature of 10K. 

\begin{figure}
    \centering
    \includegraphics[width=\hsize]{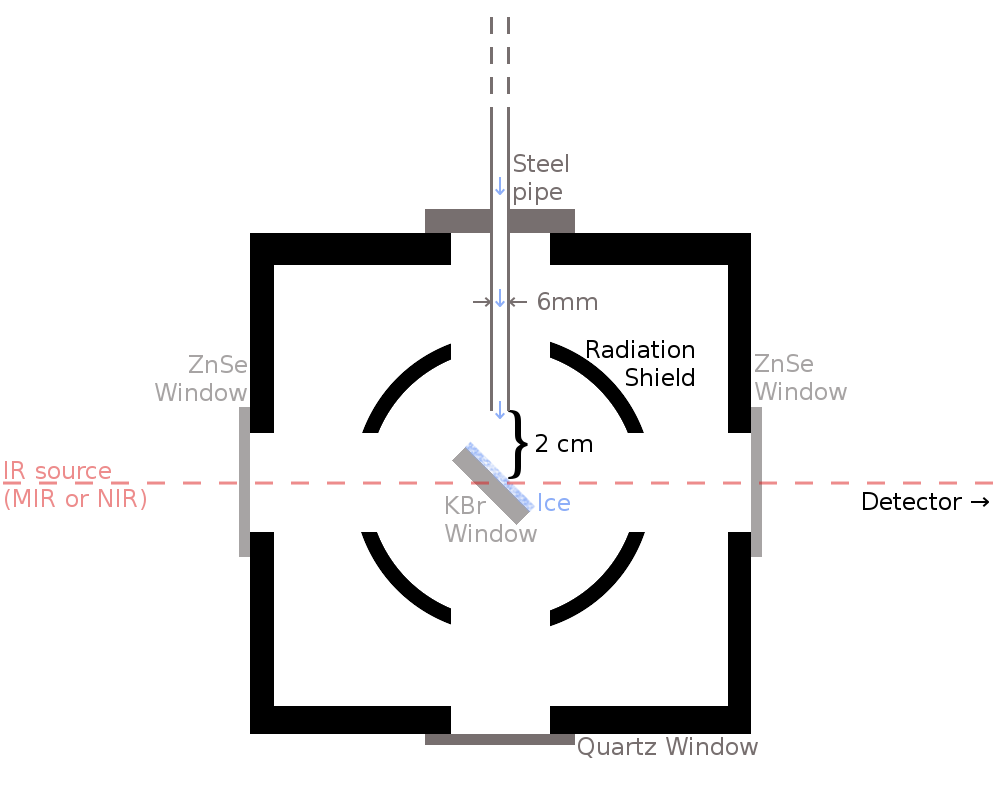}
    \caption{Sketch of the cryocooler set-up.}
    \label{fig:setup}
\end{figure}

The vacuum chamber is complemented by two ZnSe optical windows, and one quartz window for visual inspection. For the measurements in the mid-infrared (MIR) and in the near-infrared (NIR) spectral regions, with a resolution of 2 cm$^{-1}$, a \ce{DTGS} (deuterated triglycine sulphate) detector and a nitrogen cooled 
\ce{InSb} detector, respectively, were used. A globar lamp was used as radiation source.
A scheme of the arrangment of the cryocooler and its main components is illustrated in Figure \ref{fig:setup}.

The gas mixtures were prepared in a 2 liter glass flask by using standard manometric techniques. The ratios of \ce{H2O} to the \ce{O2} were adjusted by measuring the pressure inside the glass flask following the ideal gas law
\begin{equation}
 pV = nRT,
\end{equation}
whereas the number of \ce{H2O} and \ce{O2} particles is set by the isotherm relation
\begin{equation}
\frac{n_i}{n_{tot}} = \frac{p_i}{p_{tot}} = \frac{V_i}{V_{tot}}, 
\end{equation}
where $n_i$ is the particle number, $p_i$ the partial pressure and $V_i$ the partial volume of the gas \textit{i}, in our case \ce{H2O} and \ce{O2}. The total pressure $p_{tot}$ is the sum of all partial pressures 
$\sum\limits_i p_i$, $n_{tot} = \sum\limits_i n_i$ and $V_{tot} = \sum\limits_i V_i$.

The gas mixture composition used for each experiment is outlined in Table~\ref{exp}. Distilled water that had been degassed in a freeze-pump-thaw cycle three times, and oxygen with a purity of 99.999 \% were introduced in the cooling chamber under high vacuum with a base pressure of $10^{-6}$ mbar.
The deposition time was varied starting with deposition times of 30 s, using a moderate deposition rate, and increasing the deposition time up to 3 min with fast depostion rate, so that thicknesses of 0.03 $\mu$m ($\sim$ 100 MLs) 
up to 3 $\mu$m ($\sim$ 10000 MLs), with an uncertainty of 20-30 \%, could be analysed. In the fourth column of Table~\ref{exp}, the water column density for thick ice reached after 1 minute of deposition time for each ice mixture is listed; the derived water ice thickness and the oxygen column density are indicated in columns 5 and 6, respectively.
We note that the latter has a high error for water rich ices due to limits in determing the area of the weak oxygen feature that is overlapped by the water bending vibration mode. Also, a comparison of $N_{\ce{H2O}}$ and $N_{\ce{O2}}$ to estimate the real ice composition is difficult for oxygen rich ices due to monomer and dimer features in the water bands.

The thickness of the deposited ice has been calculated from the area of the three bands characteristic of the water ice spectral signature. First, the column density $N$ has been estimated from the band area for the water component, following the relation 
\begin{equation}
 N = ln(10) \cdot \frac{Area}{A},
\end{equation}
and for $A$ using the band strength values from \citet{Hagen1981} and \citet{Gerakines1995}.
The factor $ln(10)$ originates in the relation of the optical depth $\tau = Abs \cdot ln(10)$ where $Abs$ is the absorbance. $\tau$ is incorporated into the area.
Then, the thickness $d$ (cm) has been calculated assuming 
\begin{equation}
 d = \frac{N \cdot M}{\rho \cdot N_A},
\end{equation}
where $N$ is the column density (molecules\ cm$^{-2}$), $\rho$ the density (g cm$^{-3}$), $N_A$ the Avogadro constant (molecules mol$^{-1}$) and $M$ the molecular mass (g mol$^{-1}$).
The total thickness has been estimated by considering the water to oxygen molecular ratio and scaling the thickness value accordingly. For this estimation we assumed that the ice molecular composition remains constant over the deposition time scale.
There is no spectroscopic evidence which can prove that this assumption is correct, but assuming that the sticking coefficient is not varying significantly for the two components water and oxygen, we can estimate to produce a quite uniformly mixed ice. This is the general assumption used for this kind of experimental set-ups.

The cold KBr substrate is mounted with an angle of 45$^{\circ}$ with respect to the molecular beam direction. This configuration allows the deposition of the gas to take place simultaneously with the recording of the spectra.
In the vacuum chamber, the molecules expand to a collisionless beam and the ice formed on the top of the substrate has a porous structure, with an expected density of approximately 0.8 g/cm$^{-3}$ \citep{Dohnalek2003,Snels2011}.
The thickness values provided for these experiments have been scaled to the direction normal to the substrate surface, to correct for the 45$^{\circ}$ angle of the substrate with respect to the spectroscopic path.

The measurements in our experimental setup work in transmittance and thus we would take an ice deposition on both sides of the substrate into account. Nevertheless, we experimentally confirm that ice was deposited only on the side of the substrate that faces the steel pipe.
We used the laser interference method to ensure that no significant amount of ice is deposited on the substrate face opposite to the gas inlet.

\begin{table*}
\caption{Gas mixture composition, expressed as partial pressure ratio, the water column density ($N_{\ce{H2O}}$), the water ice thickness $d_{\ce{H2O}}$ and \ce{O2} column density ($N_{\ce{O2}}$)
after a total deposition time of 1 min for thick ice for each experiment presented in this paper. The uncertainty for $N_{\ce{H2O}}$ and $d_{\ce{H2O}}$ lie within 20-30 \%.}
\noindent\makebox[\textwidth]{
\begin{tabular}{l c c c c c}
\hline \hline
Exp & \ce{H2O} & \ce{O2} & $N_{\ce{H2O}}$/10$^{18}$ mol cm$^{-2}$ & $d_{\ce{H2O}}$ ($\mu$m) & $N_{\ce{O2}}$/10$^{18}$ mol cm$^{-2}$\\ [0.5ex]
\hline
\#1 & 9 & 1 & 1.94 & 0.68 & 0.02\\
\#2 & 4 & 1 & 1.42 & 0.50 & 0.02\\
\#3 & 3 & 1 & 1.12 & 0.39 & 0.07\\
\#4 & 2 & 1 & 2.17 & 0.77 & 0.21\\
\#5 & 1 & 1 & 0.95 & 0.33 & 0.28\\
\#6 & 1 & 2 & 0.58 & 0.21 & 0.16\\
\#7 & 1 & 4 & 0.25 & 0.09 & 0.32\\
\#8 & 1 & 9 & 0.08 & 0.03 & 0.25\\
\hline
\end{tabular}
}
\label{exp}
\end{table*}

\section{Results}

\subsection{\ce{O2} features in thin and thick ice samples}

\begin{figure*}
    \centering
    \includegraphics[width=\textwidth]{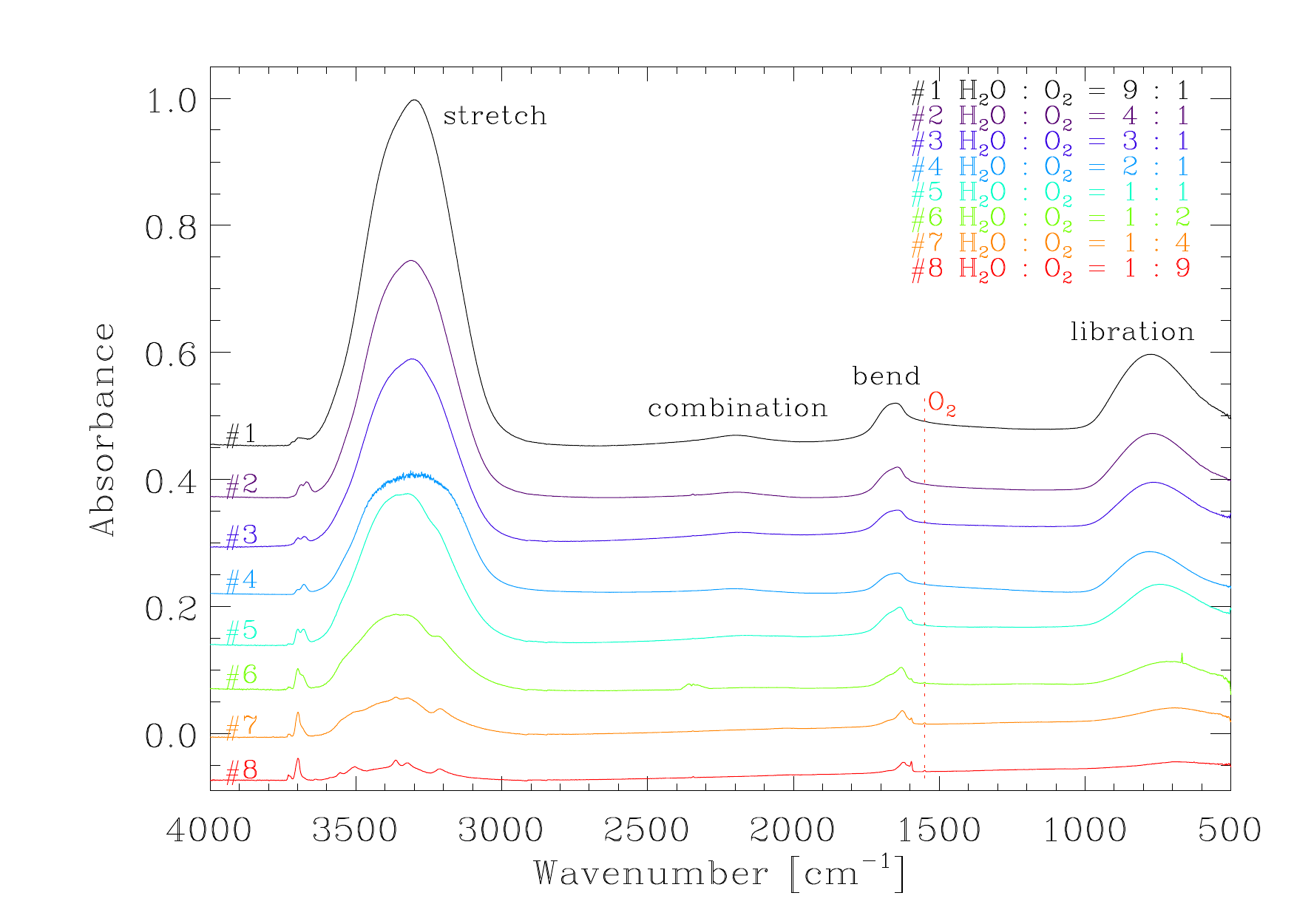}
    \caption{Spectra of all \ce{H2O}-\ce{O2} composition ratios for thick ice with same deposition time of 1 min each. The number of \ce{H2O} monolayers here goes from 80 ML (thickness $\approx$ 0.03 $\mu$m) for \ce{H2O}:\ce{O2} = 1:9 to 1900 ML (thickness $\approx$ 0.68 $\mu$m) for \ce{H2O}:\ce{O2} = 9:1. 
    The spectra for the different compositions are shifted in the absorbance scale for helping their visualization. The vertical red dashed line indicates the \ce{O2} band position at 1551 cm$^{-1}$.}
    \label{fig:spectrum_thick_labeled}
\end{figure*}

Figure \ref{fig:spectrum_thick_labeled} shows the whole observed spectrum for the different composition ratios after same deposition time.
The figure clearly shows the \ce{H2O} streching mode at 3280 cm$^{-1}$, the \ce{H2O} combination mode at 2200 cm$^{-1}$, the \ce{H2O} bending mode at 1660 cm$^{-1}$ and the \ce{H2O} libration mode at 760 cm$^{-1}$.
Moreover, the \ce{O2} feature can be found at 1551 cm$^{-1}$. In spectra \#2, \#6, \#7, and \#8, features due to \ce{CO2} contamination are present at 2343 cm$^{-1}$ and at 660 cm$^{-1}$.
The \ce{H2O}:\ce{O2} = 2:1 spectrum shows the saturation of the stretching mode because of different deposition conditions compared to the other experiments.

\begin{figure*}
    \centering
    \includegraphics[width=\textwidth]{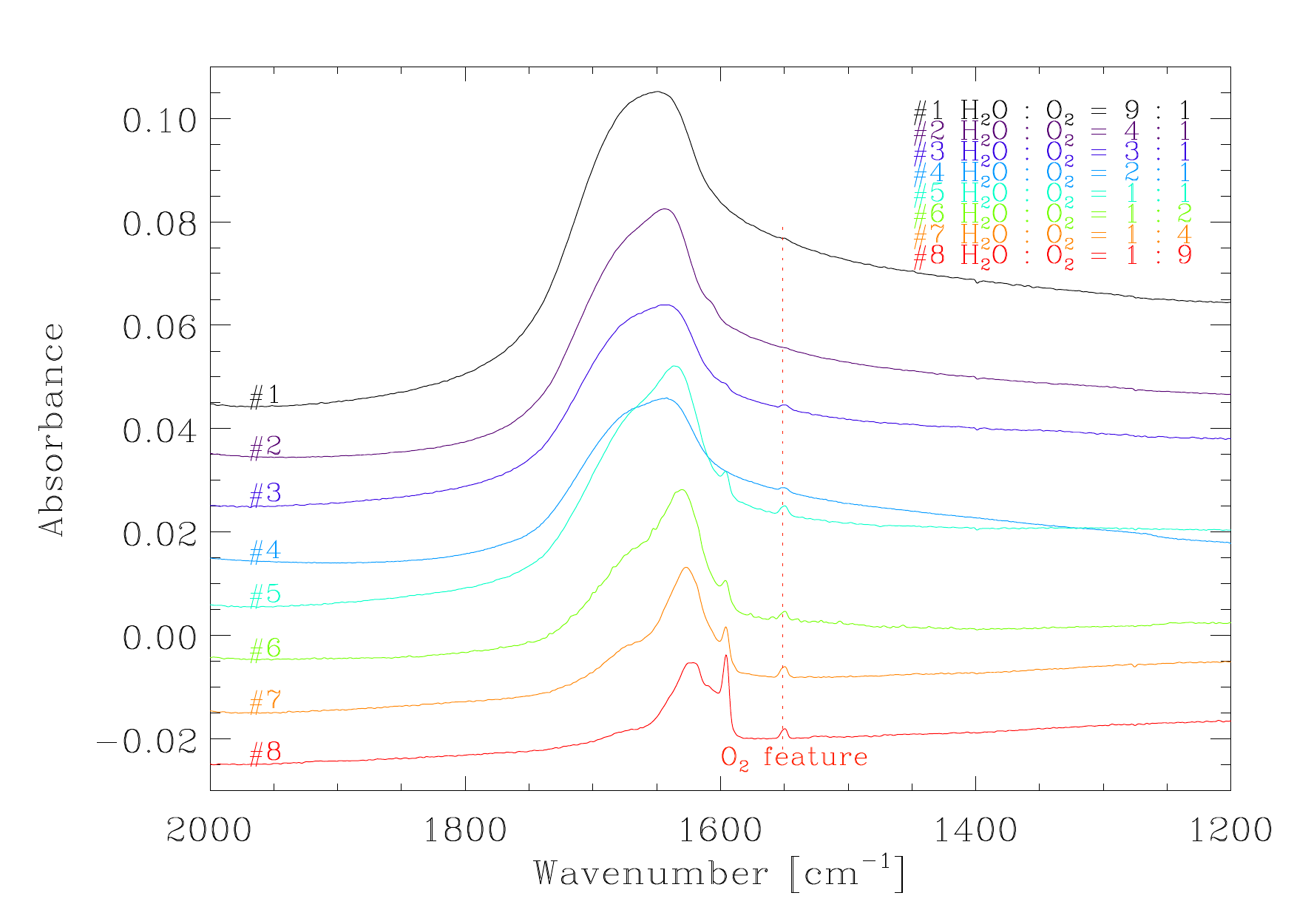}
    \caption{Magnification of spectra of Fig. \ref{fig:spectrum_thick_labeled} around the \ce{O2} feature (marked by the vertical red dashed line) for all \ce{H2O}-\ce{O2} mixtures for thick ice.}
    \label{fig:O2_all_thick}
\end{figure*}

\begin{figure}[!ht]
    \centering
    \includegraphics[width=\hsize]{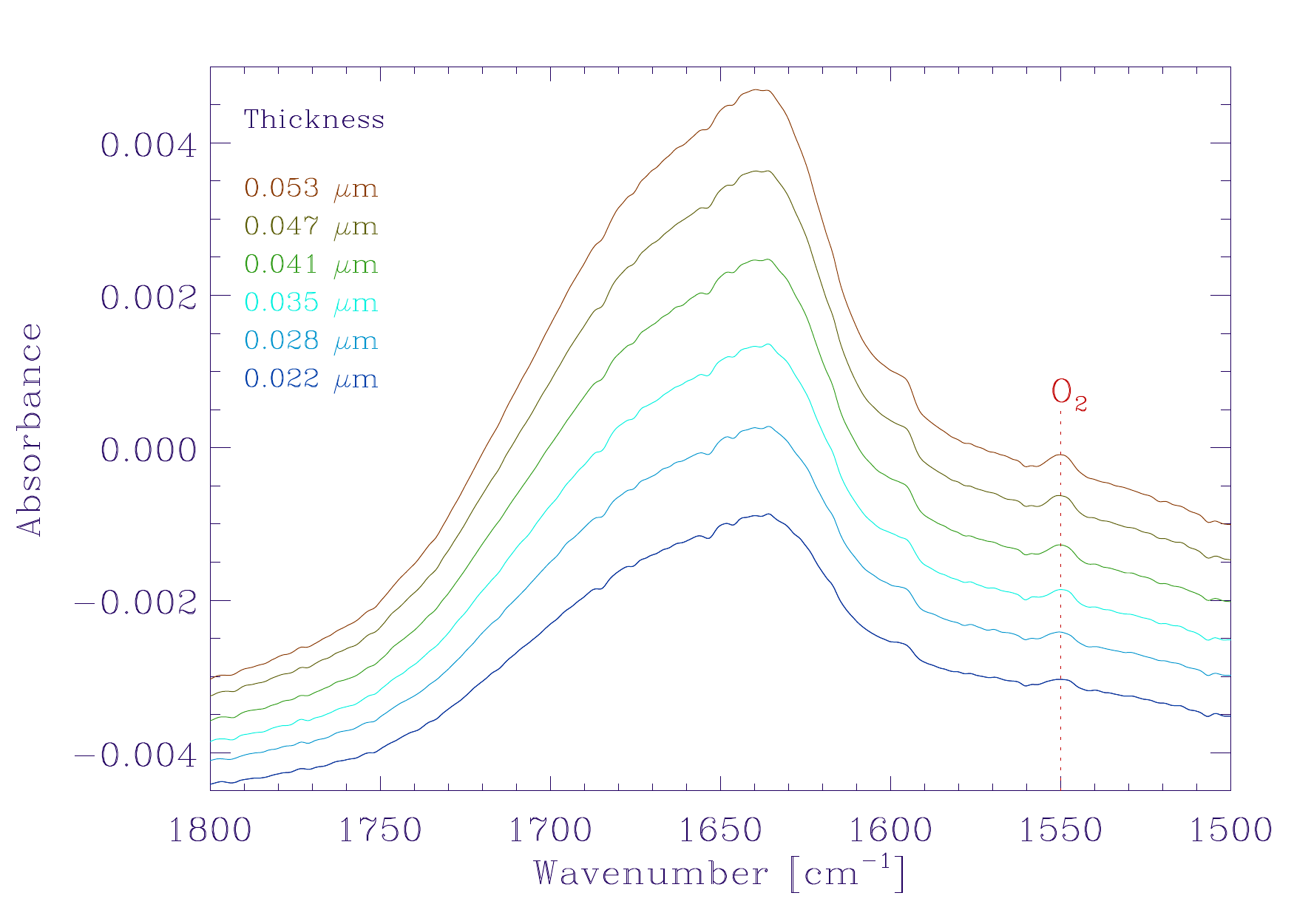}
    \caption{Magnified spectra of \ce{H2O}:\ce{O2} = 2:1 mixture around the \ce{O2} feature; the ice thicknesses is between 0.022 and 0.053 $\mu$m for thin ice.}
    \label{fig:O2_thin_new}
\end{figure}

\begin{figure}[!ht]
    \centering
    \includegraphics[width=\hsize]{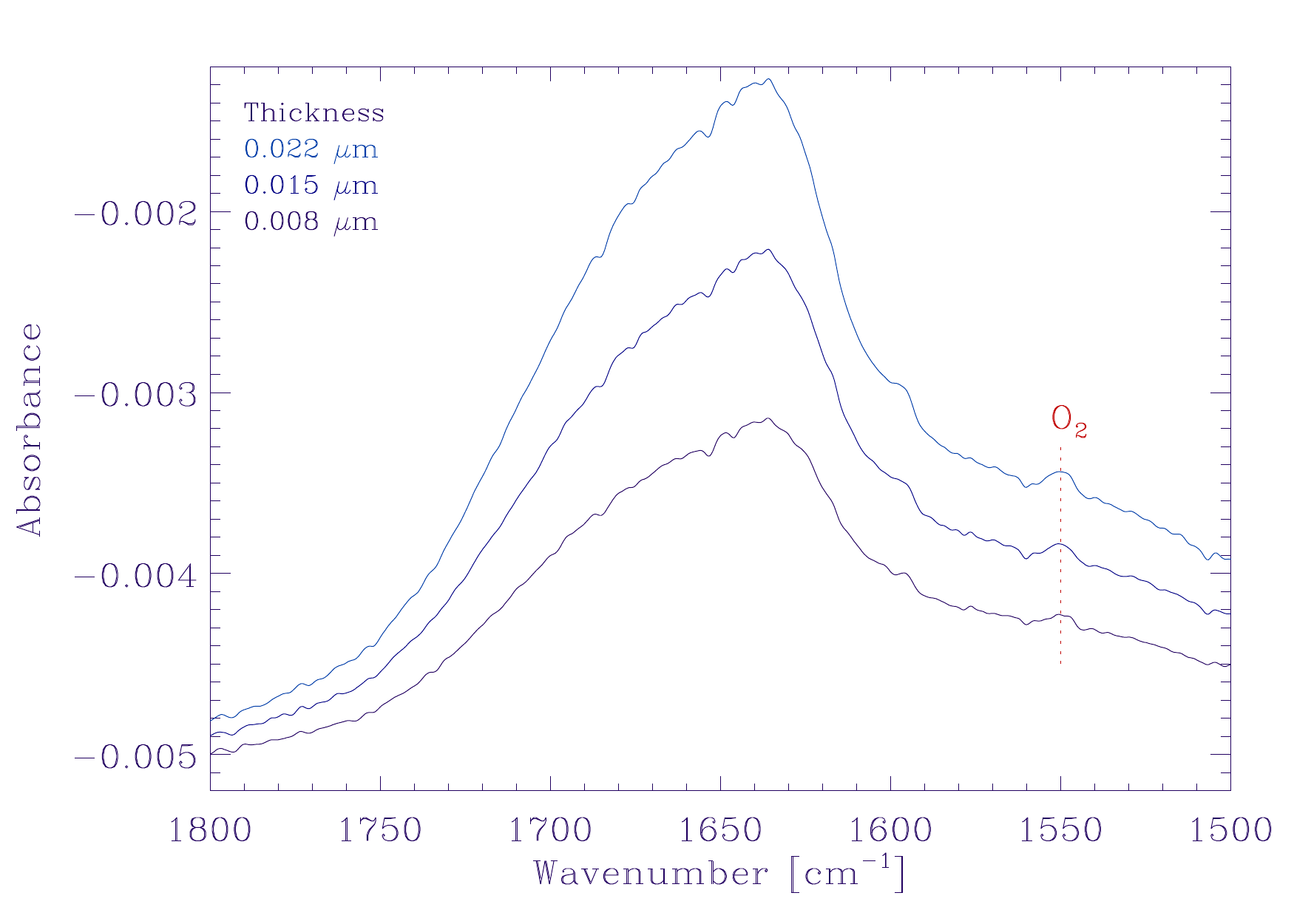}
    \caption{Magnified spectra of \ce{H2O}:\ce{O2} = 2:1 mixture around the \ce{O2} feature; the thicknesses is between 0.008 and 0.022 $\mu$m for thin ice.}
    \label{fig:O2_thin_new_10-30s}
\end{figure}

In the thick ice layer regime (estimated thicknesses ranging between 1 and 3 $\mu$m) the oxygen feature is visible at all the investigated molecular ratio. Our limit cases are the mixture in excess of water (experiment \#1) and the mixture in excess of 
oxygen (experiment \#8). It is convenient to divide the analysis into two parts, one for the experiments from \#1 to \#4, and a second for the experiments from \#5 to \#8. In the first part, excess of water at different molecular ratios is investigated. 
In these experiments, the oxygen feature is overlapped to the bending vibration mode of water, making its detection increasingly difficult as the water becomes predominant in the molecular ratio (cf. Figure \ref{fig:O2_all_thick}). Nevertheless, even with a water to oxygen ratio of 9:1, the oxygen 
feature is still visible in our recorded spectrum, as it shown in Figure \ref{fig:O2_all_thick}.

For the thin ice, we conducted separate experiments in order to monitor the dependence of the \ce{O2} band position on the physical conditions of the thin ice.
The spectrum of the thin \ce{H2O}:\ce{O2} = 2:1 ice (thickness 0.022 - 0.053 $\mu$m) is shown enclosed around 1660 cm$^{-1}$ in Figure \ref{fig:O2_thin_new}.
Presenting even thinner ice, Figure \ref{fig:O2_thin_new_10-30s} shows the first three deposition steps (thickness 0.008 - 0.022 $\mu$m) in order to enhance the visibility of the \ce{O2} feature of which we see hints of the band at 1551 cm$^{-1}$ also for the thinnest ice.
The water bands assume the characteristic features of water trapped in a matrix, as can be observed from \citet{Ehrenfreund1996}.

In the experiments in excess of oxygen, the \ce{O2} feature is progressively isolated from the water bending feature. Its spectral characteristics, though, are not varying significantly with
the change in the \ce{H2O} to \ce{O2} ratio, in terms of frequency shift and band profile. Therefore, these spectra offer the best conditions for the estimate of the \ce{O2} band strength.
The band strength $A$ is given as
\begin{equation}
 A = \int \sigma(\tilde{\nu}) d\tilde{\nu} = \frac{1}{N} \int \tau(\tilde{\nu}) d\tilde{\nu},
\end{equation}
where $\sigma(\tilde{\nu})$ is the cross-section (cm$^2$) and $\tau(\tilde{\nu})$ the optical depth for a given wavenumber $\tilde{\nu}$ (cm$^{-1}$).
Using
\begin{equation}
 \tau = Abs \cdot ln(10),
\end{equation}
where $Abs$ is the measured absorbance of the \ce{O2} feature, we get
\begin{equation}
 \int \tau(\tilde{\nu}) d\tilde{\nu} \approx Abs \cdot ln(10)\ d\tilde{\nu}.
\end{equation}
We measured $d\tilde{\nu}$ = 10 cm$^{-1}$ and for \ce{O2} the column density $N = 3.01 \times 10^{22}$ molecules cm$^{-3} \cdot d.$
The derived band strengths for the different ice mixtures and a comparison with the values found by \citet{Ehrenfreund1992} can be found in Table~\ref{band_strength}.

\begin{table*}
\centering
\begin{threeparttable}
\caption{Band strength (A) of \ce{O2} for the different ice compositions.}
\label{band_strength}
\begin{tabular}{c c | l l c}
\hline \hline
\ce{H2O} : \ce{O2} & A\tnote{a} (cm\ molec$^{-1}$) & Mixture\tnote{b} & Ratio\tnote{b} & A\tnote{b} (cm\ molec$^{-1}$) \\
\hline
9 : 1 & 7.6e-21 & & & \\
4 : 1 & 9.6e-21 & & & \\
3 : 1 & 1.5e-20 & & & \\
2 : 1 & 4.8e-20 & \ce{H2O} : \ce{CO} : \ce{CO2} : \ce{O2} & 2 : 2 : 0.5 : 1 & 1.0e-19\\
1 : 1 & 3.9e-20 & \ce{H2O} : \ce{CO} : \ce{O2} & 1 : 1 : 1 & 0.7e-19\\
1 : 2 & 6.4e-20 & & \\
1 : 4 & 1.6e-19 & & \\
1 : 9 & 3.2e-19 & & \\
& & \ce{CO} : \ce{O2} & 1 : 1 & 0.7e-19\\
& & \ce{CO2} : \ce{O2} & 1 : 1 & 6.0e-19\\
& & \ce{CO2} : \ce{O2} & 10 : 1 & 3.0e-18\\
\hline
\end{tabular}
\begin{tablenotes}
\item [a] The values have been averaged for the analysis of different ice thicknesses.
\item [b] Comparison with band strength as derived by \citet{Ehrenfreund1992}. \ce{H2O} to \ce{O2} ratios with equal values to those used in our work have been aligned to facilitate comparison.
\end{tablenotes}
\end{threeparttable}
\end{table*}

For comparable experiments, where the values of \ce{H2O}/\ce{O2}  presented in this work are equal to those presented by \citet{Ehrenfreund1992}, the values of the band strength agree with a factor of approximately two.
This difference is most likely due to the influence of \ce{CO} and \ce{CO2} in the experiments conducted by \citet{Ehrenfreund1992}, which enhance the oxygen band strength.

As the feature of solid \ce{O2} is present at 1551 cm$^{-1}$, there exists the possibility for observable overtones in the NIR. While we don't expect to see any \ce{O2} features near 3102 cm$^{-1}$ (3.22 $\mu$m) due to the strong water signal, 
we looked for the next possible overtone at 4653 cm$^{-1}$ (2.15 $\mu$m). Even, for the thick ice with 3 $\mu$m thickness, there is no \ce{O2} feature to be found near 4653 cm$^{-1}$.

\subsection{Dangling bonds}

We note that in our experiments we dealt with a porous amorphous ice structure. For this reason, in the recorded spectra we observe the features assigned to the presence of the so-called dangling bonds, that is, features arising from the OH stretching 
vibration of water molecules that are not engaged in a intermolecular interaction with the ice bulk. Although the analysis of the spectral signature of the dangling bonds is not directly applicable to the comparison with astronomical observations, 
for which no detection is currently available, we believe that studying the way in which the presence of oxygen affects the features ascribed to the dangling bonds can provide useful information for future laboratory experiments.
In Figure \ref{fig:dangl_scaled} the spectral region of the dangling bonds {for thick ice corresponding to the different \ce{H2O}-\ce{O2} mixtures and the pure water is shown.
The increasing steepness of the spectra towards 3500 cm$^{-1}$ relate to the increasing ratio of \ce{H2O} to \ce{O2} as the stretch feature of water rises in intensity with increasing water abundance.
The positions of the dangling bond features can be found in Table \ref{table:dangbond}.

\begin{figure*}
    \centering
    \includegraphics[width=\textwidth]{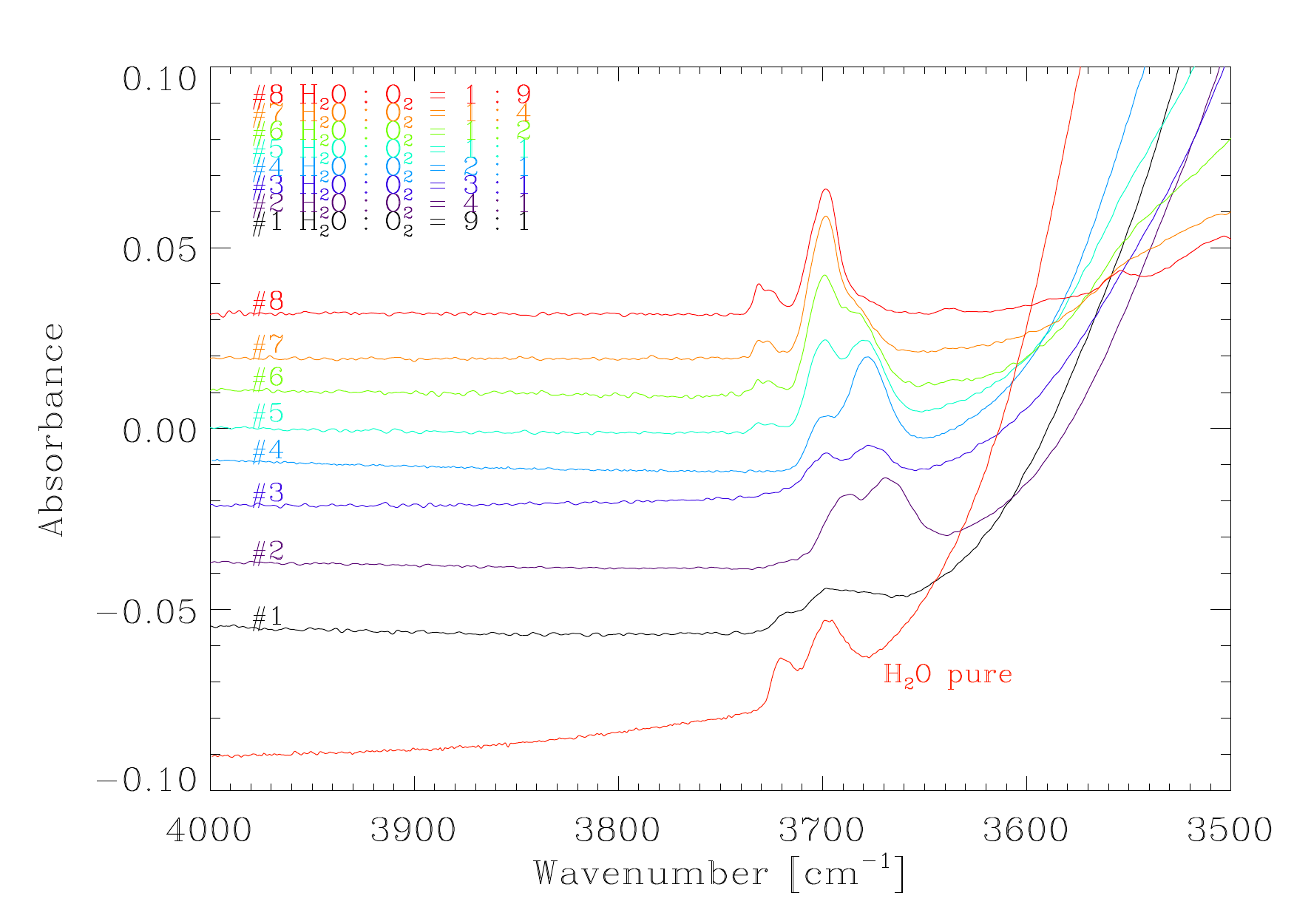}
    \caption{Thick-ice absorption spectrum of \ce{H2O}:\ce{O2} mixtures compared to pure water. To enable a better overview, similar to Fig. \ref{fig:spectrum_thick_labeled}, the spectra of the different mixtures have been shifted.}
    \label{fig:dangl_scaled}
\end{figure*}

\begin{table}
\caption{Gas mixture composition and position of the dangling bonds (with shoulder feature position where present) for each experiment presented in this paper.}
\centering
\begin{tabular}{l c c c}
\hline \hline
Exp & \ce{H2O} : \ce{O2} & Dang. bond \#1        & Dang. bond \#2 \\
    &                    & (shoulder)/cm$^{-1}$  & (shoulder)/cm$^{-1}$ \\
\hline
\#1 & 9 : 1 & 3717 & 3698 \\
\#2 & 4 : 1 & 3689 & 3669 \\
\#3 & 3 : 1 & 3698 & 3678 \\
\#4 & 2 : 1 & 3693 & 3678 \\
\#5 & 1 : 1 & 3731 (3725) & 3699 (3680) \\
\#6 & 1 : 2 & 3731 (3727) & 3699 (3686) \\
\#7 & 1 : 4 & 3731 (3727) & 3699 (3685) \\
\#8 & 1 : 9 & 3731 (3725) & 3699 (3680) \\
\hline
\end{tabular}
\label{table:dangbond}
\end{table}

Figure \ref{fig:dangl_scaled} shows a red shift of the dangling bonds peaks going from the pure water to the \ce{H2O}:\ce{O2} = 2:1 mixture, from 3697 cm$^{-1}$ (2.705 $\mu$m) and 3683 cm$^{-1}$ (2.715 $\mu$m) to 3698 cm$^{-1}$ (2.704 $\mu$m) and 3678 cm$^{-1}$ (2.719 $\mu$m).
The change in the spectroscopic feature of the dangling bonds for the water and oxygen ices with respect to pure water ice is in agreement with previous experiments performed by \citet{Palumbo2010}, in which the effect on the profile of the dangling bond 
feature for an ice mixture of approximately 20 \% of oxygen in water is reported. Their study can be compared to our \ce{H2O}:\ce{O2} = 4:1 mixture. 
For oxygen-rich ice mixtures, the peak shape and position is probably influenced by the presence of water monomers and dimers embedded in the oxygen matrix, as reported in \citet{Ehrenfreund1996}.

\section{Discussion}

We decided to limit our study to the case of the molecular oxygen complexes with water and to extend the amount of molecular ratio explored going from ice mixtures in excess of water to excess of oxygen, in order to have a complete overview of the molecular oxygen signature in different astrophysical environments.
Furthermore, for each experiment we have analysed the IR spectra in the thin and thick ice layer regimes, showing that the oxygen spectral signature does not exhibit different characteristics, in terms of band position and shape, with respect to the two approaches.}
The molecular vibration of \ce{O2} by itself is inactive in the infrared since the dipole moment doesn't change upon molecular vibration.
Nevertheless, the interaction with the water matrix breaks the electron symmetry and induce a small dipole moment, which allows for the detection in the IR region.

In the thin ice layer regime, in which the thicknesses of the ice layer (estimated for the water component) varys between 0.008 and 0.053 $\mu$m, the oxygen feature has been observed only in the most favourable case, that is, the \ce{H2O}:\ce{O2} = 2:1 
mixture (experiment \#4). In this molecular ratio, the band strength of the oxygen feature is enhanced, as shown by \citet{Ehrenfreund1992}, the instrumental detection limit in dealing with a low column density is overcome, and 
the observation of a low intensity band is possible.
For even thinner ices however, we don't expect the shape and position of the \ce{O2} feature to change much, but the detection of \ce{O2} depends on the total column density, so in the end it depends on the ice thickness as well as dust number density.
In the remaining cases the low intensity of the band signal did not allow for a detection.
The frequency position measured in the experiment \#4 is 1551 cm$^{-1}$ (6.45 $\mu$m).
In order to increase the S$/$N for the thin \ce{H2O}:\ce{O2} = 2:1 ice, the recording of the spectra was performed with a higher number of recorded cycles (2048 record cycles; for comparison: 1184 record cycles for thick ice).
That way, the noise was reduced in order to obtain a S$/$N $\lesssim$ 3 even for the thinnest ice.

In addition, we compared the dangling bond features of pure water with that of the \ce{H2O}:\ce{O2} = 9:1 mixtures.
As already observed by \citet{Palumbo2010}, the profile of the dangling bond feature depends on the presence of other species mixed-in with water ice.
The presence of oxygen affects the dangling bonds profile in term of intensity and shape. Their profile may be related to the microstructure of the deposited ice.
As can be observed from Figure \ref{fig:dangl_scaled}, the \ce{H2O}:\ce{O2} ratio in the ice is affecting significantly the spectral characteristic of the water dangling bonds.
Going from the water rich to oxygen rich mixtures, we observe a progressive change in the relative intensity of the two main features and a shift in their position.
For oxygen rich mixtures, we observe a predominant sharper feature at lower frequencies, while the water rich mixtures exhibit a smoother profile and similar relative intensities of the two components.
The mixture with a similar water to oxygen ratio shows a profile which is in between the extremes.
From this consideration we can relate the shape of the dangling bonds bands with the molecular composition of the bulk ice, allowing an identification of the enrichment of the ice in one or the other species.
However, a detailed analysis of the effect of the structure of the ice on the dangling bond characteristic is outside the scope of this work.

\begin{figure}
    \centering
    \includegraphics[width=\hsize]{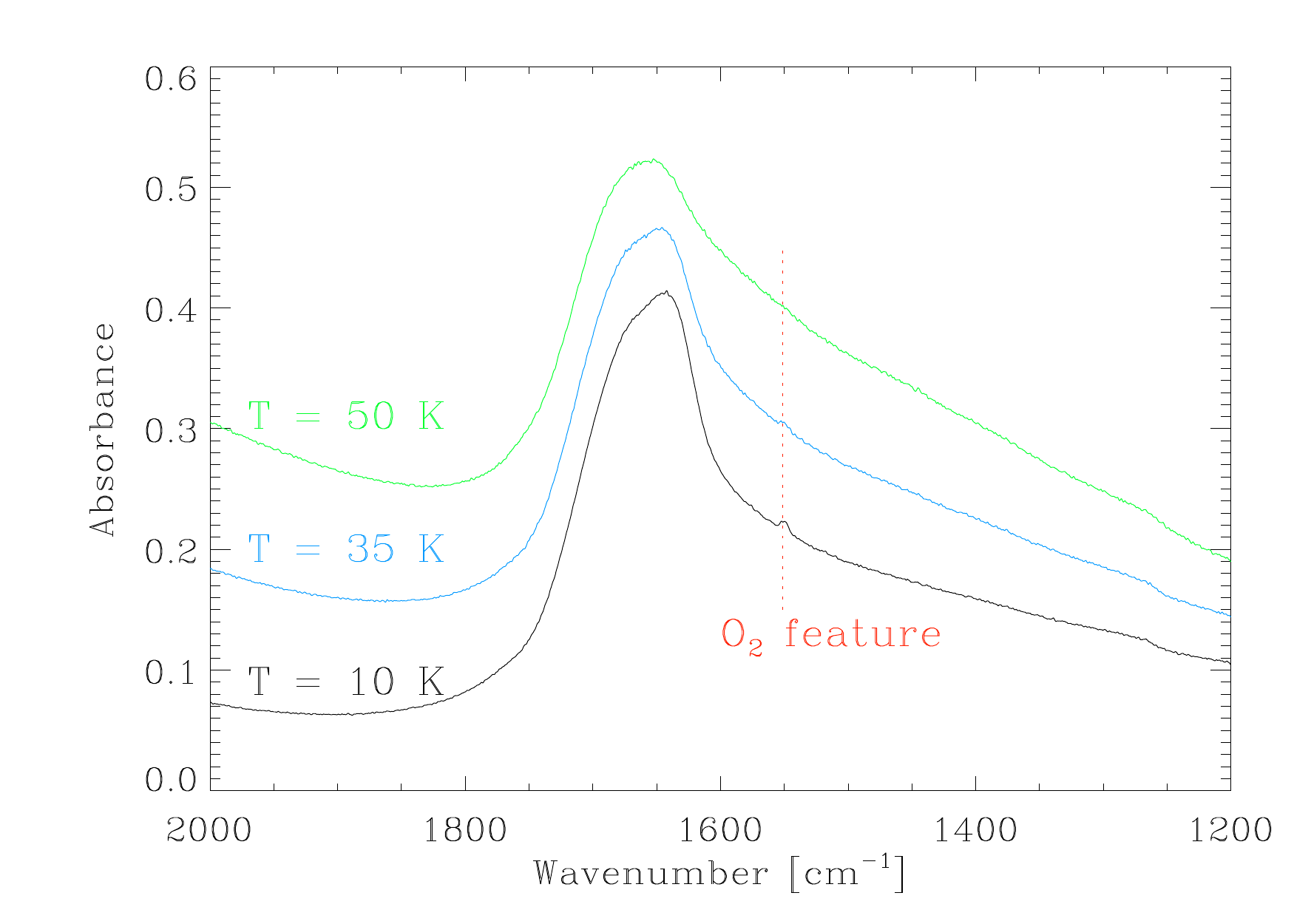}
    \caption{Behaviour of the \ce{O2} feature (marked by the vertical dashed line) for \ce{H2O}:\ce{O2} = 2:1 while heating the ice from 10 K to 50 K.}
    \label{fig:O2_heating}
\end{figure}

Figure \ref{fig:O2_heating} shows the behaviour of the \ce{O2} feature upon heating.
Observed \ce{O2} features started to vanish when the probes were heated to 35 $\mathrm{K}$.
At this temperature, the \ce{O2} molecules trapped on the amorphous water ice matrix start to evaporate \citep[cf.][]{Bar-Nun1987}.
At a temperature of 50 $\mathrm{K}$, the \ce{O2} features were completely absent.
Above this temperature the \ce{O2} is not trapped by the porous amorphous solid water any more and the \ce{O2} is desorbed from the deposited ice.
This observation is consistent with the experiments conducted by \citet{Ayotte2001}.

\section{Astrophysical implications}

Our experimental approach intends to explore astrophysical scenarios. The experiments dedicated to the thin ice layer regime (number of monolayers $<$ 150; counted using the thickness and density of the ice) simulate the conditions in the dense interstellar medium more adequately \citep{Oeberg2016},
while the experiments in the thick layer regime attempt to extend the physical conditions relative to the centre of pre-stellar cores as well as in the midplane of protoplanetary disks, where pebbles and icy bodies precursors of comets and planetesimals with ice thicknesses $>$ 200 ML form \cite[e.g.][]{Bieler2015}. There is a lack of observational data proving the presence of molecular oxygen in ISM ice, while \ce{O2} has been recently detected in the coma of the comet 67P/Churyumov-Gerasimenko 
\citep{Bieler2015, Mousis2016}. 

Our experimental data show that the detection of the \ce{O2} feature in thin ices is much more difficult compared to the thick ones, and the only molecular composition allowing the laboratory detection is the \ce{H2O}:\ce{O2} = 2:1, which is unrealistic for 
the ISM ice composition.
However, predictions by astrochemical models for pre-stellar cores, such as L1544, indicate that the icy mantles can reach thicknesses close to the thick ice regime in these experiments \citep[e.g.][]{Vasyunin2017}.
On the other hand, the mixture with a molecular ratio \ce{H2O}:\ce{O2} = 9:1 will account for a realistic proportion like it was observed in the cometary coma of 67P/Churyumov-Gerasimenko with a mean value of 3.80 $\pm$ 0.85 \%.

\subsection{Chemical modelling} \label{Chemical modelling}

We studied which ice structure and composition is predicted by current astrochemical models, and how it is related to the observational data on \ce{O2}:\ce{H2O} ratio and ice thickness.
For this purpose, we used results from \citet{Vasyunin2017}. Briefly, this is a three-phase (gas-reactive ice surface - (less) reactive ice bulk) time-dependent astrochemical model that includes an extensive set of gas-phase and grain-surface chemistry.
In \citet{Vasyunin2017}, the model has been successfully applied to the explanation of observed distribution of complex organic molecules (COMs) in the prototypical pre-stellar core L1544. The authors find that the best agreement between the model and the observed data is reached after 1.6 $\times$ 10$^{5}$ years of chemical evolution of a static cloud with physical structure similar to one reproduced for L1544.
The model also reproduced the enchanced abundances of COMs at the so-called methanol peak located $\sim$4000 AU ($\sim$0.015 pc) away from the centre of the core (the peak of the dust emission), as observed by \citet{Bizzocchi2014}.

\begin{figure*}
    \centering
    \includegraphics[width=0.9\textwidth]{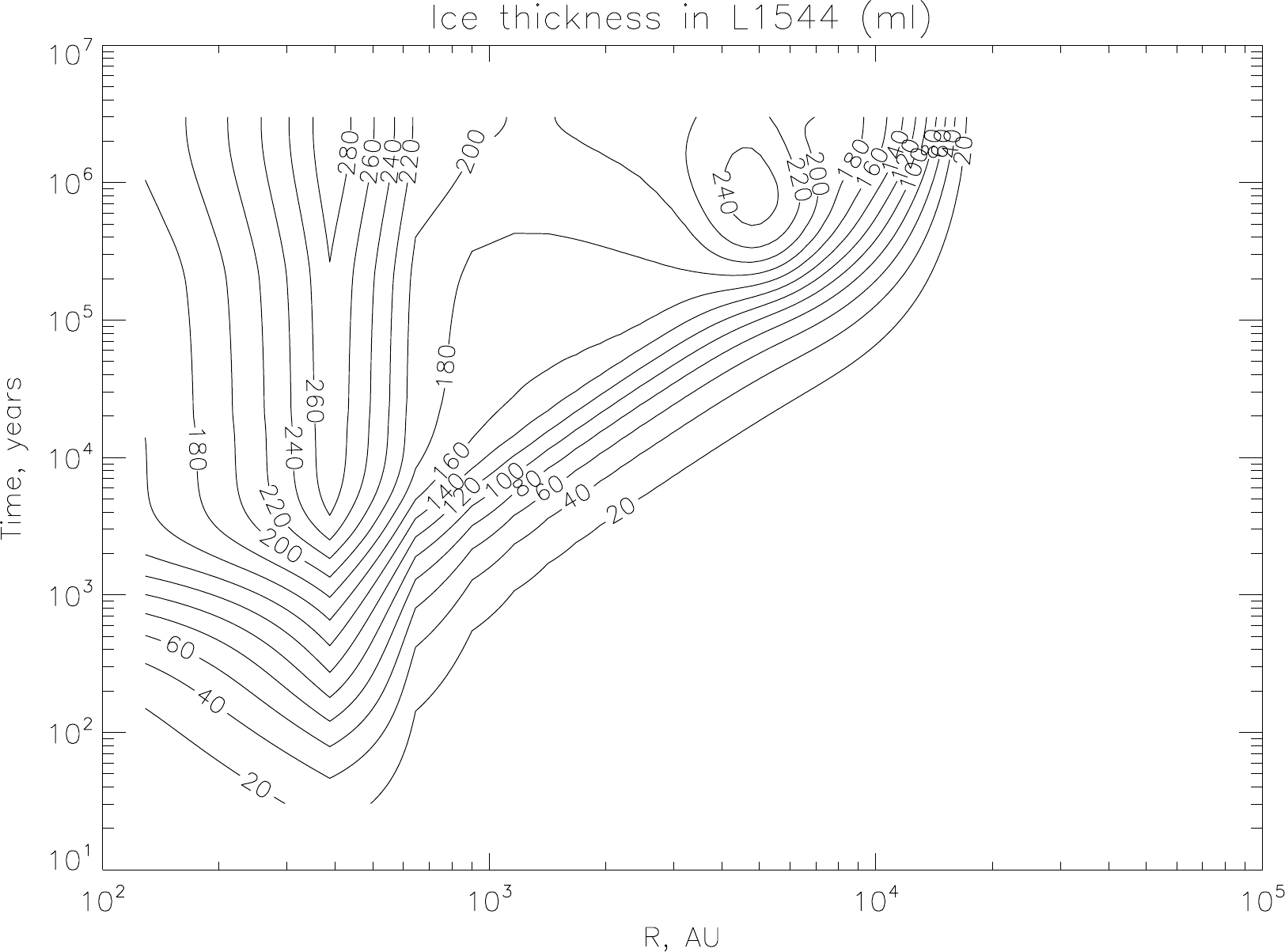}
    \caption{Model ice thickness (in monolayers) vs. time in L1544. The `methanol peak' is located 4000 AU away from the core centre.}
    \label{fig:ice_th_ml}
\end{figure*}

At the time of maximum \ce{O2}/\ce{H2O} abundance, the modelled ice thickness is 70 monolayers, but it is growing with time, and stabilizes at the methanol peak at 200 monolayers after 10$^{5}$ years of evolution, as shown in Figure \ref{fig:ice_th_ml}.
As such, we believe that modelling results are in rough agreement with observed values for \ce{O2}/\ce{H2O} in the coma of 67P/Churyumov-Gerasimenko comet.
Also, the modelling suggests that icy mantles in pre-stellar cores such as L1544 are relatively thick, and may reach several hundreds of monolayers.
Finally, it is interesting to note that while the model by \citet{Vasyunin2017} predicts gas-phase \ce{O2} abundance, at the time of the best agreement between the model and observed values of COMs, to be at least an order of magnitue lower than that of \ce{CO}, the \ce{O2} gas-phase abundance is still overestimated by an order of magnitude in comparison to observed upper limits in other environments \citep[e.g.][]{Goldsmith2000}.
The gas-phase \ce{O2} abundance in the model by \citet{Vasyunin2017} is mainly controlled by the reactive desorption mechanism, whose efficiency under various conditions is currently a matter of debate \citep[see e.g.][]{Minissale2016,Chuang2018,He2017,Oba2018}. Thus, gas-phase abundance of \ce{O2} will be the subject of further theoretical and experimental studies.

\subsection{Astronomical observability}

\begin{figure}
    \centering
    \includegraphics[width=\hsize]{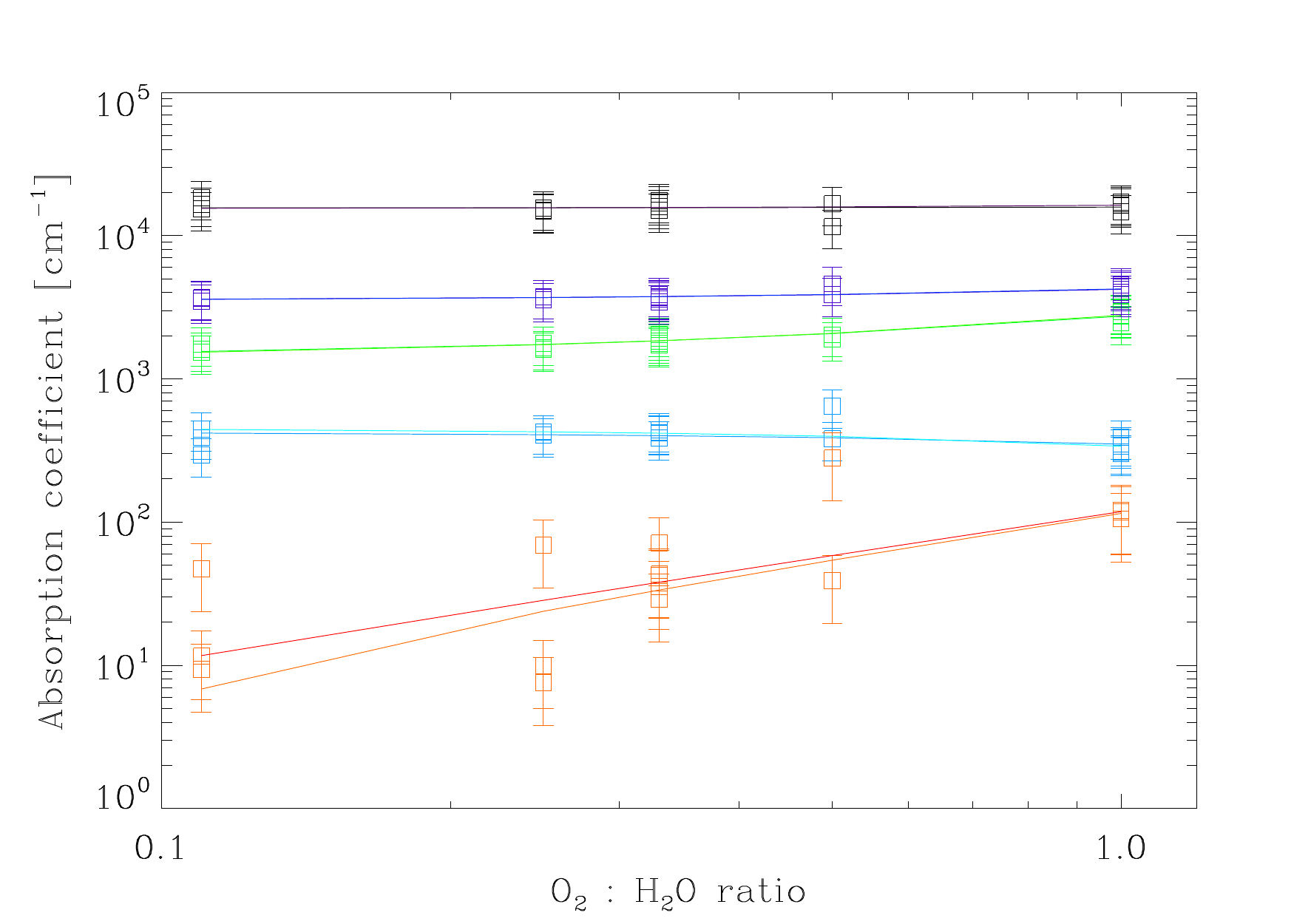}
    \caption{Absorption coefficients of the 3280 cm$^{-1}$ (black), 2200 cm$^{-1}$ (light blue), 1660 cm$^{-1}$ (green), 760 cm$^{-1}$ (dark blue) water and 1551 cm$^{-1}$ (red) \ce{O2} bands for the different mixture ratios of \ce{O2} and \ce{H2O}.
    The data were fit using the a) minimizing $\chi^2$-error statistics (bright coloured fits) and b) `robust' least absolute deviation method (dark coloured fits).}
    \label{fig:abs_coeff_ratio}
\end{figure}

Using absorbance and thickness data obtained in our experiments, we were able to calculate the absorption coefficient
\begin{equation}
 \alpha = \frac{Abs \cdot ln(10)}{d},
\end{equation}
for the different mixture ratios of \ce{O2}/\ce{H2O}.
$\alpha$ as a function of the \ce{O2}/\ce{H2O} ratio can be found in Figure \ref{fig:abs_coeff_ratio}. The absorption coefficients for water are in good agreement with the derived values from \citet{Warren2008}, considering a discrepancy in sample temperature, while we have no possibility to compare the calculated absorption coefficient for \ce{O2} with the literature.
In order to calculate $\alpha$ for all \ce{O2} to \ce{H2O} ratios, the data were fit using two different methods for statistical comparison.
The first method creates the fit by minimizing the $\chi^2$-error statistics. The second method uses the least absolute deviation method that minimizes the sum of absolute errors.

\begin{figure}[!ht]
    \centering
    \includegraphics[width=\hsize]{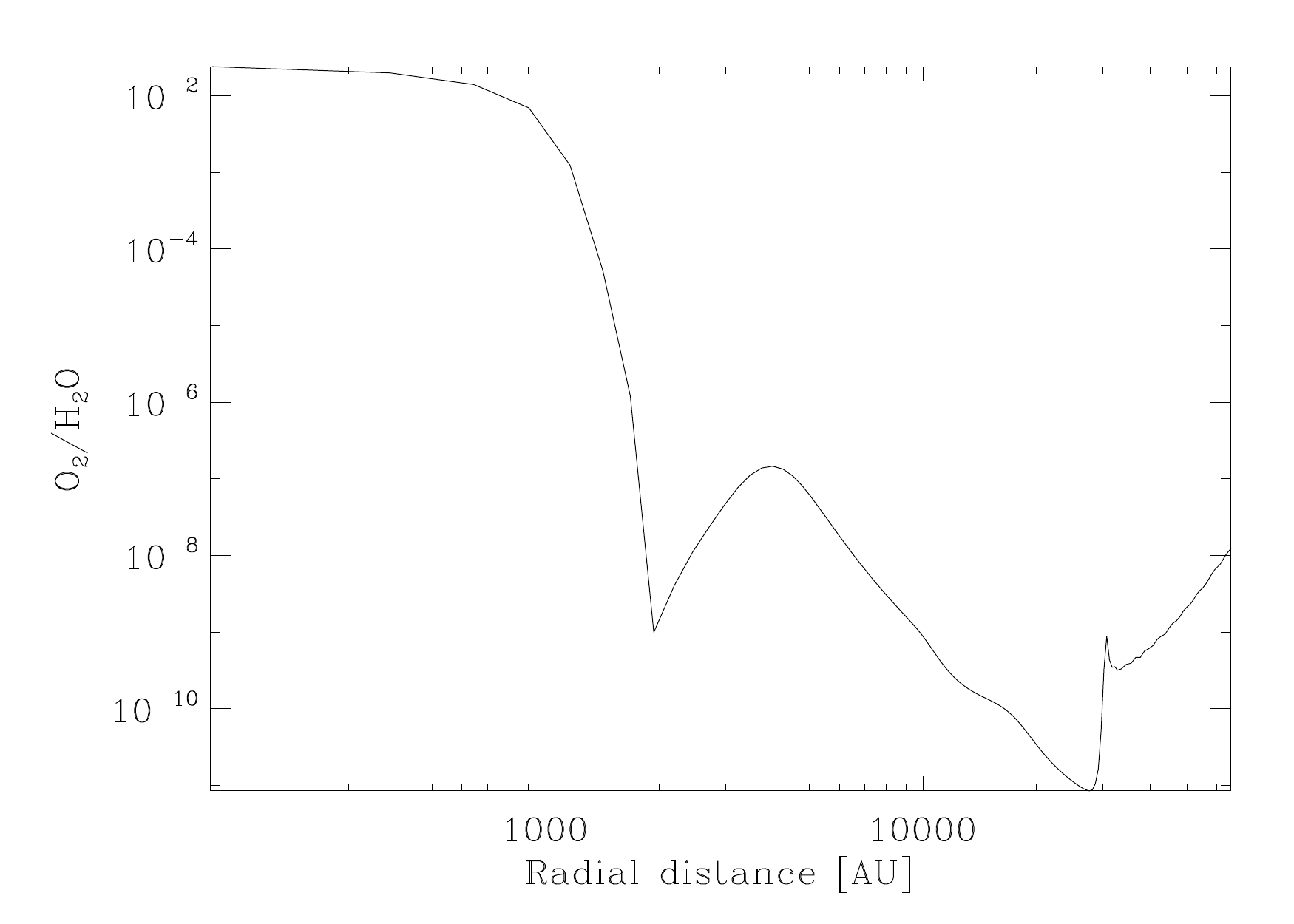}
    \caption{Calculated \ce{O2}/\ce{H2O} of L1544 as a function of the radial distance for t = 1.6 $\times 10^5$ yr.}
    \label{fig:Anton_ratio_dist}
\end{figure}

The calculated values were then used to derive the optical depth and the transmittance of L1544 using the model by \citet{Vasyunin2017} introduced in section \ref{Chemical modelling} (cf. Figure \ref{fig:Anton_ratio_dist}).
The transmittance of an area with a constant absorption coefficient can be calculated with
\begin{equation}
 T = \frac{F}{F_0} = e^{- \int \alpha s} = e^{- \tau},
\end{equation}
where $F_0$ is the initial flux density received from the medium, $F$ is the flux density transmitted by the medium, $s$ is the size of the area with a certain absorption coefficient $\alpha$.
For our calculations, we divide the modelled cloud into subpaths $s_n$ of constant absorption coefficients and obtain
\begin{equation}
 \tau_{tot} =  \sum_{s_n} \tau_{s_n}.
\end{equation}
Following this, we get the total transmittance
\begin{equation}
 T_{tot} = \prod_{s_n} T_{s_n}.
\end{equation}
The optical depth and transmittance was computed only for the inner part of our modelled L1544 where \ce{O2}/\ce{H2O} is $>10^{-6}$.
The derived optical depth and transmittance for the different \ce{H2O} and \ce{O2} can be found in Table \ref{table:L1544_transm}.

\begin{table*}
\caption{Optical depth and transmittance of the modelled L1544 in the area with \ce{O2}/\ce{H2O} $>10^{-6}$.}
\noindent\makebox[\textwidth]{
\begin{tabular}{l c c c c}
\hline \hline
Band & $\tau$ (fitting method a) & $\tau$ (fitting method b) & T (fitting method a) & T (fitting method b) \\
\hline
3280 cm$^{-1}$ & 0.625 $\pm$ 0.006 & 0.611 $\pm$ 0.039 & 0.535 $\pm$ 0.005 & 0.543 $\pm$ 0.039 \\
2200 cm$^{-1}$ & 0.017 $\pm$ 0.001 & 0.018 $\pm$ 0.002 & 0.982 $\pm$ 0.052 & 0.982 $\pm$ 0.093 \\
1660 cm$^{-1}$ & 0.056 $\pm$ 0.002 & 0.055 $\pm$ 0.003 & 0.945 $\pm$ 0.024 & 0.946 $\pm$ 0.048 \\
1551 cm$^{-1}$ & 7.4 $\times 10^{-4} \pm$ 8.1 $\times 10^{-4}$ & 1.0 $\times 10^{-5}$ $\pm$ 3.0 $\times 10^{-6}$ & 0.999 $\pm$ 0.110 & 1.000 $\pm$ 0.337 \\
760 cm$^{-1}$ & 0.140 $\pm$ 0.003 & 0.139 $\pm$ 0.006 & 0.869 $\pm$ 0.016 & 0.870 $\pm$ 0.035 \\
\hline
\hline
\end{tabular}
}
\label{table:L1544_transm}
\end{table*}

Directly comparing the overlapping \ce{H2O} 1660 cm$^{-1}$ and \ce{O2} 1551 cm$^{-1}$ bands we get a transmittance ratio of 0.946 for fitting method a) and b).
The predicted transmittance of the \ce{O2} band is then $\approx 5.4 \%$ of the water band transmittance at 1660 cm$^{-1}$. In order to check the observability of this feature with the \textit{James Webb Space Telescope} (JWST) the time calculator facility has been employed.
For the prestellar core L1544, we calculated the time needed to detect the \ce{O2} line at a 3$\sigma$ level for the MIRI instrument in the Channel 1 (short) configuration and found that this requires more than 100 hours with JWST, meaning it is not feasible.
This estimate is, however, conservative, as we expect, based on the experiments by \citet{Ehrenfreund1992}, that the \ce{O2} feature becomes stronger when ice mixtures are considered.
We will quantify this in a future paper, where we will present the spectroscopic signatures of an ice mixture similar to the one predicted by \citet{Vasyunin2017}.

\section{Conclusions}
This paper presents spectral features of solid molecular oxygen enclosed in a water matrix. The analysis is compared to previous studies reported by \citet{Ehrenfreund1992} and \citet{Palumbo2010}, and extended to 
further molecular ratios and ice thicknesses, as motivated by the recent detection of \ce{O2} in the comet 67P/Churyumov-Gerasimenko.
In particular, we have explored the possibility of detecting \ce{O2} in solid form towards pre-stellar cores for which gas-grain chemical model predictions were available.

In our experiments, \ce{O2} is found at 1551 cm$^{-1}$ for thick ice. Also, we find evidence for the \ce{O2} band at that position even for our thinnest ice.
Because of their more heterogeneous composition, interstellar ices may, however, show the \ce{O2} features around these values.
The porous amorphous nature of the ice mixture produces the dangling bonds near the stretching mode of water.
However, while their position and intensity depends on the ratio of \ce{H2O} and \ce{O2}, the lack of detection in space hint to the conclusion that the ice there is compact and not porous \citep[e.g.][]{Palumbo2005}.
The \ce{O2} feature can only be seen for temperatures $\le$ 35 K.

The behaviour of the dangling bonds in dependence of the ice mixture composition has been discussed. The \ce{O2}:\ce{H2O} ratio not only has an effect on the intensity of the features but also on their position and shape.
This is especially so for mixtures with an excess of \ce{O2} that show a shoulder feature or even a second maximum near the two main bands.
  
The three-phase time-dependent astrochemical model of the pre-stellar core L1544, as introduced by \citet{Vasyunin2017} predicts thick ice ($\approx$ 200 ML) and \ce{O2}/\ce{H2O} fractions close to those measured towards the comet 67P.
Using JWST, it is not feasible to detect \ce{O2} in solid phase via measuring the transmission ratio of L1544 using the MIRI instrument. New estimates for the \ce{O2} detection in mixed ices with JWST, however, will be done in a future paper.

\section*{Acknowledgements}
The authors thank the anonymous referee, who greatly helped us to enhance the clarity of the paper.
We thank Christian Deysenroth for the very valuable contribution in the designing and mechanical work of the experimental setup.
We are also grateful to Miwa Goto, Wing-Fai Thi, and Seyit Hocuk for fruitful discussion.
The work of Birgitta M\"uller was supported by the IMPRS on Astrophysics at the Ludwig-Maximilians University, as well as by the funding the IMPRS receives from the Max-Planck Society.
The work of Anton Vasyunin is supported by the Russian Science Foundation via the Project No. 18-12-00351.

\end{document}